\documentclass[aps,apl,reprint,superscriptaddress,floatfix]{revtex4-1}
\usepackage{color,units}
\usepackage{amssymb}
\usepackage{amsmath}
\usepackage{psfrag}
\usepackage{topcapt}
\usepackage{hyperref}
\usepackage{float}
\usepackage{braket,color,soul}
\usepackage[normalem]{ulem}

\usepackage[utf8]{inputenc}

\usepackage{ifpdf}

\ifpdf
\usepackage[pdftex]{graphicx}
\else
\usepackage{graphicx}
\fi

\begin{document}
\title{High-Kinetic Inductance Additive Manufactured Superconducting Microwave Cavity}

\author{Eric T. Holland}
\email{holland25@llnl.gov}
\affiliation{Lawrence Livermore National Laboratory, Livermore, CA 94550 USA}

\author{Yaniv J. Rosen}
\affiliation{Lawrence Livermore National Laboratory, Livermore, CA 94550 USA}

\author{Nicholas Materise}
\affiliation{Lawrence Livermore National Laboratory, Livermore, CA 94550 USA}

\author{Nathan Woollett}
\affiliation{Lawrence Livermore National Laboratory, Livermore, CA 94550 USA}

\author{Thomas Voisin}
\affiliation{Lawrence Livermore National Laboratory, Livermore, CA 94550 USA}

\author{Y. Morris Wang}
\affiliation{Lawrence Livermore National Laboratory, Livermore, CA 94550 USA}

\author{Jorge Mireles}
\affiliation{The University of Texas at El Paso,W.M. Keck Center, El Paso, TX 79968 USA}

\author{Gianpaolo Carosi}
\affiliation{Lawrence Livermore National Laboratory, Livermore, CA 94550 USA}

\author{Jonathan L DuBois}
\affiliation{Lawrence Livermore National Laboratory, Livermore, CA 94550 USA}

\date{\today}

\ifpdf
\DeclareGraphicsExtensions{.pdf, .jpg, .tif}
\else
\DeclareGraphicsExtensions{.eps, .jpg}
\fi

\begin{abstract}
Investigations into the microwave surface impedance of superconducting resonators {have led to the development of single photon counters that rely on kinetic inductance for their operation}.  {While concurrent progress in additive manufacturing, `3D printing', opens up a previously inaccessible design space for waveguide resonators.} In this manuscript, we present results from the first synthesis of these two technologies in a titanium, aluminum, vanadium (Ti-6Al-4V) superconducting radio frequency resonator which exploits a design unattainable through conventional fabrication means. We find that Ti-6Al-4V has two distinct superconducting transition temperatures observable in heat capacity measurements.  The higher transition temperature is in agreement with DC resistance measurements.  While the lower transition temperature, not previously known in literature, is consistent with the observed temperature dependence of the superconducting microwave surface impedance. From the surface reactance, we extract a London penetration depth of $8 \pm 3\mu m$ -- roughly an order of magnitude larger than other titanium alloys and several orders of magnitude larger than other conventional elemental superconductors. This large London penetration depth suggests that Ti-6Al-4V may be a suitable material for high kinetic inductance applications such as single photon counting or parametric amplification used in quantum computing.\end{abstract}

\maketitle
The Landau-Ginzburg model for superconductivity is a mean field theory that parametrizes the superconducting condensate with two key length scales \cite{landau1950,gor1959}.  The first length scale is the Pippard coherence length, $\zeta$, which is the average distance between Cooper pairs in the condensate.  The second length scale, the London penetration depth, $\lambda$, describes the distance over which a magnetic field penetrates into a superconductor.  The ratio between the Pippard coherence length and the London penetration depth is commonly used to characterize whether a superconductor is type I ($\zeta/\lambda>\sqrt{\frac{1}{2}}$), field expelling, or type II ($\zeta/\lambda<\sqrt{\frac{1}{2}}$), field penetrating material \cite{tinkham1996}.  As such, precise measurements of either the Pippard coherence length or, as in the case of this manuscript, the London penetration depth are critical for the accurate description of the superconducting material.

In superconducting radio frequency resonators, the London penetration depth describes the distance over which energy is stored in the motion of Cooper pairs.  This energy storage mechanism is associated with a current and therefore described as a kinetic inductance.  Kinetic inductance is determined, in part, by the equilibrium temperature in relation to the superconducting transition temperature.  This feature of kinetic inductance has been exploited in a number of applications including, for example, radiation detectors with up to single photon resolution by so called microwave kinetic inductance detectors (MKIDs) \cite{day2003}.  

In this Letter, we show that titanium, aluminum, vanadium, Ti-6Al-4V, (90\% Ti, 6\% Al, 4\% V by weight) is a high kinetic inductance superconductor, that the hexagonal close packed (alpha) phase is indeed a superconductor contrary to suspicion in previous literature \cite{leyens2003}, and that additive manufacturing (AM) enables a new parameter space of seamless resonator designs left unexplored in previous work \cite{creedon2016}.  We investigate the superconducting properties of bulk AM Ti-6Al-4V by DC resistance and heat capacity measurements.  Furthermore, we probe the microwave surface impedance by studying both the reactive, on resonance, and resistive, off resonance, behavior of an AM Ti-6Al-4V resonator at milliKelvin temperatures.  To our knowledge, this result is the first measurement of any kind of the London penetration depth for the most widely used titanium alloy in industry \cite{leyens2003}.  
\begin{figure}
\centering
\includegraphics[scale=.18]{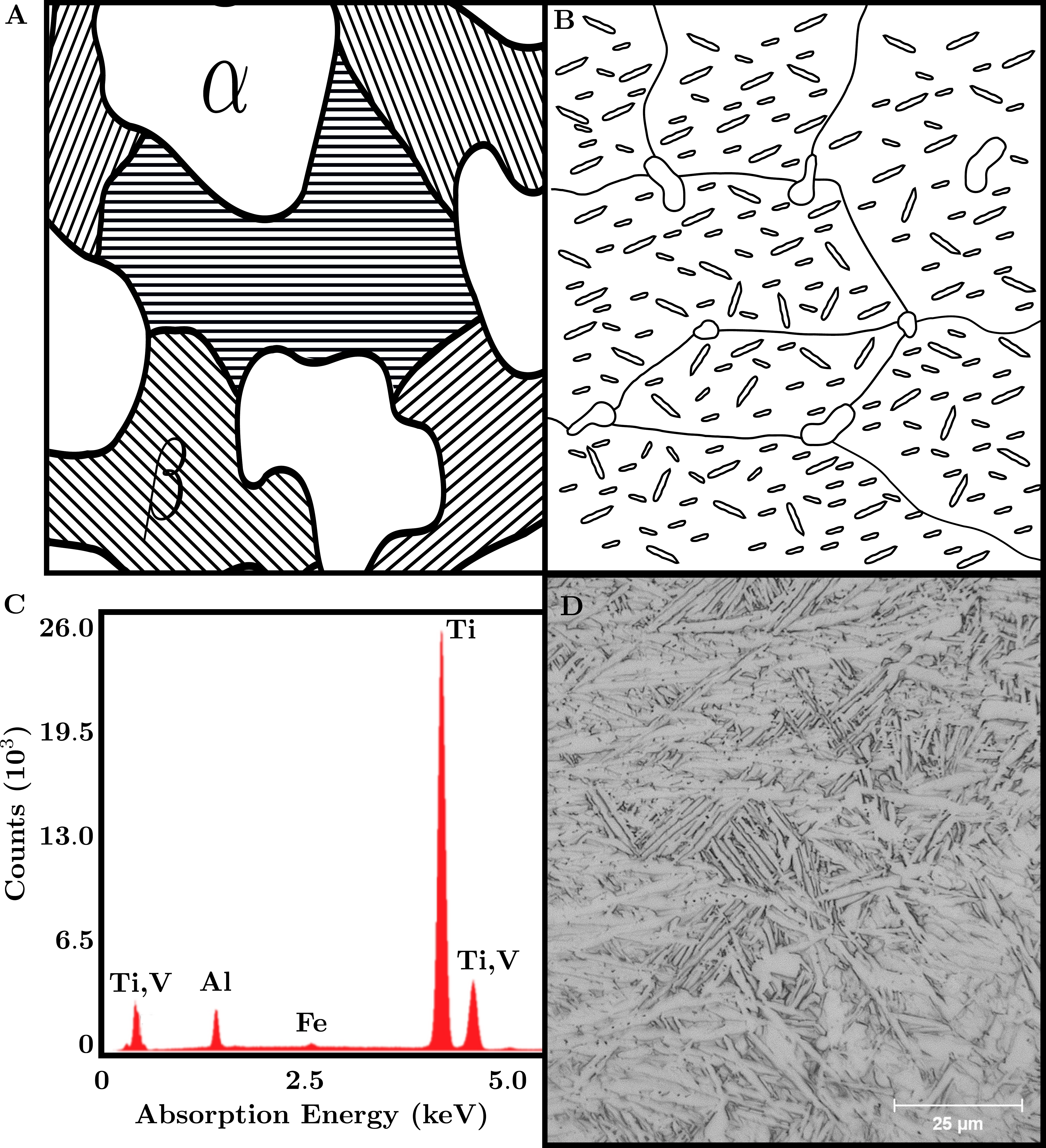}
\caption{Ti-6Al-4V Microstructure and Elemental Composition.  \textbf{A} Nominal sketch of the typical bulk microstructure of Ti-6Al-4V before temperature anneal \cite{gorman2001}.  White regions depict the hexagonal close packed, $\alpha$ phase.  Repeating line regions are body centered cubic, $\beta$ phase. \textbf{B}  Sketch of the idealized microstructure of Ti-6Al-4V for optimal mechanical performance post high temperature anneal \cite{gorman2001}.  Now white regions represent $\alpha$ phase dispersed within martensitic $\beta$ phase regions.  \textbf{C} Representative spectroscopy of the Ti-6Al-4V surface composition after acid polish.  Titanium (90\%), aluminum (6\%), vanadium (4\%), and iron ($<$.5\%) are all in percentages as used commercially for bulk Ti-6Al-4V.  \textbf{D} Scanning electron microscope image of Ti-6Al-4V microstructure after optimized acid polish.}
{\label{fig:surface}}
\end{figure}

Titanium alloys, and Ti-6Al-4V in particular, are commonly used in AM processes (see references \cite{kobryn2001,thijs2010,pyka2012} for examples).  In the present work, we have employed a selective laser melting (SLM) process \cite{kruth2007} to fabricate our samples.  SLM works by using a high powered laser to melt and solidify a metallic powder according to a digitally designed geometry.  The laser power and rastering parameters used to fabricate the resonator structures under study were optimized for mechanical performance and minimal porosity \cite{matthews2016}.  In addition, our samples received a post-processing heat treatment designed for selecting the bulk microstructure.  The heat treatment involved a half hour duration heat ramp to $900^{\circ}$C in atmosphere, which was maintained for an hour, and finally quench cooled to room temperature with argon gas.

We begin our characterization of Ti-6Al-4V by investigating the bulk microstructure and elemental composition to compare the AM sample to bulk Ti-6Al-4V.   To investigate the bulk microstructure, we performed a cross sectional cut on the sample after performing characterization of the superconducting microwave surface impedance presented later in this work.  After the cross section, the sample received an acid polish (immersed in 10 mL HF, 5 mL HNO$_3$, and 85 mL H$_2$O  for 15 seconds) to reliably produced contrast between the $\alpha$ and $\beta$ phases.  After acid polish, shown in Fig. \ref{fig:surface} \textbf{D}, is a representative scanning electron microscope (SEM) image from our sample.  Qualitatively, the surface microstructure is as expected--specifically, the sample is dense and has the expected microstructure (Fig. \ref{fig:surface} \textbf{B}).  In Fig. \ref{fig:surface} \textbf{C}, the results of energy-dispersive X-ray spectroscopy (EDS) demonstrate an elemental composition of titanium (90\%), aluminum (6\%), vanadium (4\%), and a less then percent contribution from other impurities such as iron.  Further localized EDS and SEM studies in the interior and near the sample surface were consistent with those shown with no evidence for spatial inhomogeneities in the phase, crystal structure or composition.  We therefore conclude that our samples are qualitatively equivalent to bulk Ti-6Al-4V, which has undergone a similar heat treatment during cooling, and that the results presented in this manuscript are not limited to AM Ti-6Al-4V.

\begin{figure}
\centering
\includegraphics[scale=.2]{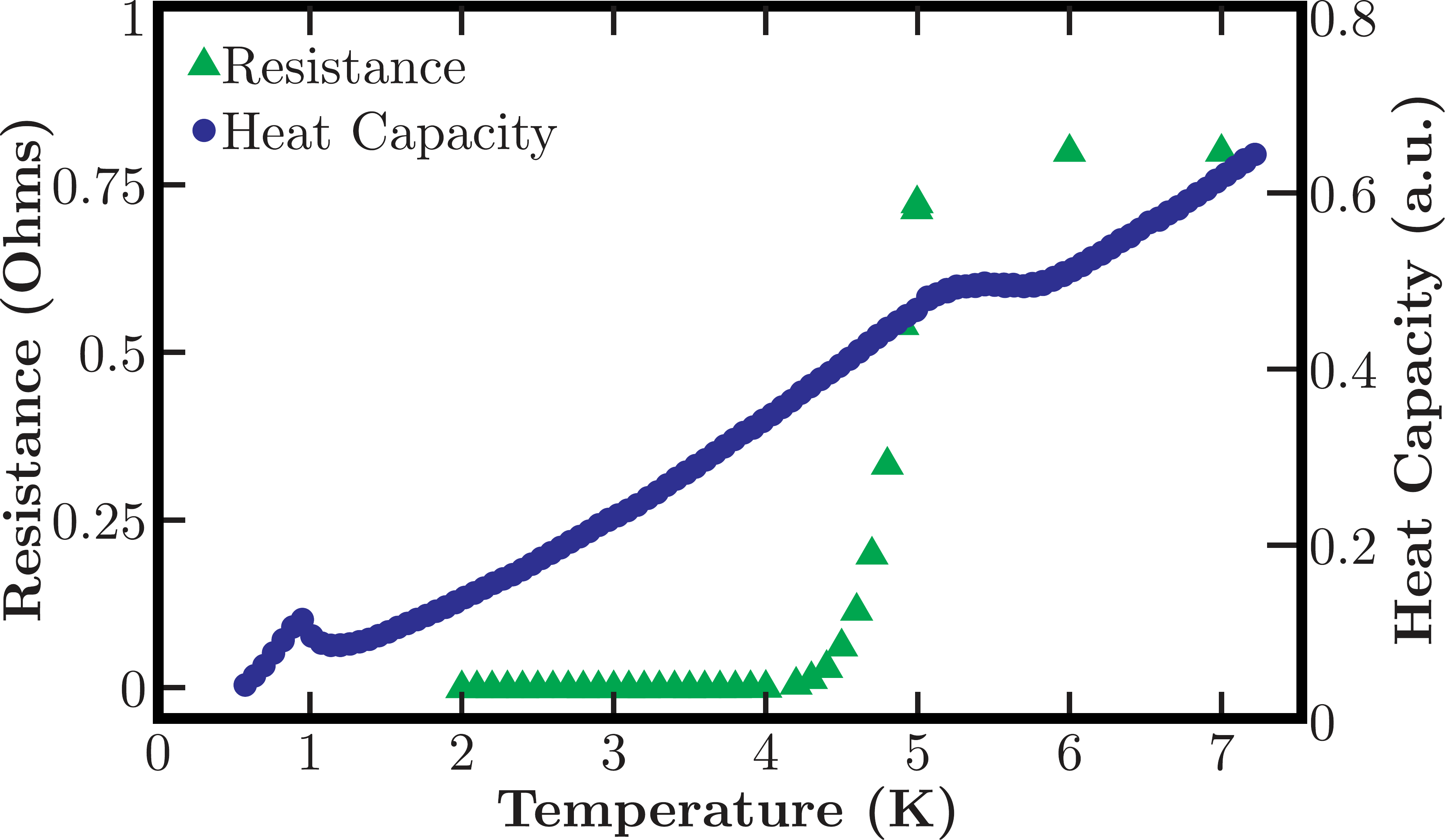}
\caption{DC Resistance and Heat Capacity.  DC resistance (left y-axis) plotted as a function of equilibrium temperature (green triangles) of a four point measurement performed in a Quantum Design Physical Property Measurement System shows a superconducting transition temperature of $4.5$ K.  Heat capacity (right y-axis) plotted as a function of equilibrium temperature (blue dots) displays two distinct superconducting phase transitions.  The lower transition temperature, 0.95 K, is not observed in the DC resistance measurements presumably because it is shorted out by other superconducting pathways.}
{\label{fig:dcheat}}
\end{figure}

Through previous DC resistance measurements, it is known that Ti-6Al-4V is a superconductor with a single transition temperatures that varies in the range of 1.3 K to 6.3 K depending on oxidation treatment as well as post processing annealing \cite{clark1970,lepper1972,wolff1973,umezawa1992,divakar2008,divakar2010,ridgeon2016}.  In agreement with this literature, we observe a superconducting transition temperature of 4.5 K ( Fig. \ref{fig:dcheat} solid green triangles, 90\% drop value) through a four point measurement in a Quantum Design Physical Property Measurement System.  However, by investigating the low temperature behavior of the heat capacity for Ti-6Al-4V, we observe two superconducting phase transitions as shown in, Fig. \ref{fig:dcheat} {blue dots}.  The first phase transition at higher temperatures is in qualitative agreement with our DC measurements.  The second phase transition is observed at a much lower temperature, 0.95 K.  This is the first time, to our knowledge, that heat capacity measurements have been employed to confirm the superconducting transition temperature of Ti-6Al-4V.  Additionally, the lower transition temperature is missing from the literature.

\begin{figure}
\centering
\includegraphics[scale=1]{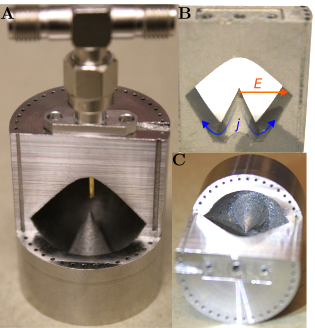}
\caption{AM Ti-6Al-4V Conical Resonator.  \textbf{A} A nominally identical cavity, is cut to reveal internal structure.  A resistive SMA T is connected to an SMA connector which is evanescently coupled to the resonant mode of the conical resonator.  \textbf{B} Cross sectional cut of the Ti-6Al-4V resonator after an acid polish.  On resonance the current flows along the base of the cone and in the walls while the electric field is maximal at the tip of the cone.  \textbf{C} Alternative angle, with the SMA connector removed, of the conical resonator showing the internal geometry and coupling mechanism.}
{\label{fig:pic}}
\end{figure}

From the above characterization, we learn that AM Ti-6Al-4V may be a robust material to fabricate superconducting resonators and proceed to characterize its superconducting microwave surface impedance in the quantum regime (i.e. $\hbar \omega <k_bT$  and the average photon number in the resonator is, $\overline{n} \approx 1$).  The results presented in this work are part of an ongoing study investigating the role of geometric effects on resonator performance.   In designing the resonator structure for this study we employed the expanded design space offered by AM fabrication to explore the relative importance of loss mechanisms arising from inductive (surface resistance) and capacitive (dielectric loss tangent) contributions.   The resonator under study in this work (see Fig. \ref{fig:pic}) is a variant of a well performing aluminum coaxial quarter wave resonator with millisecond lifetimes in the quantum regime \cite{reagor2016,wang2016} designed to minimize the surface current density on resonance.

Characterization of the superconducting microwave surface impedance was performed in a Janis model JD-250 (wet) He3-He4 dilution refrigeration system.  Cryogenic attenuation of 40 dB is included on the 1K pot stage, and 20 dB on the mixing chamber stage (20 mK base temperature) to filter the black body noise power down to the that of the mixing chamber.  The Ti-6Al-4V conical resonator and a gold plated copper box containing planar aluminum coplanar waveguide resonators were multiplexed on a high purity oxygen free copper cold finger attached to the mixing chamber.  For multiplexing, we used a resistive power dividing SMA T attached to an SMA connector with post that couples below cutoff through a cylindrical waveguide (evanescently) to the conical resonator similar to work done in reference \cite{reagor2013}.  The samples were fully enclosed in an Amuneal A4K magnetic shield attenuating ambient magnetic field to less than a milliGauss.  The magnetic shielding is thermally heat sunk to a light tight copper shield blocking stray infrared radiation.  The output of our samples led into two 4-8 GHz PAMTECH model CTH0408KI isolators in series providing reverse isolation in excess of 40 dB, allowing investigations of the resonators in the quantum regime.  An output SMA line of superconducting NbTiN led to the cryogenic HEMT amplifier (Low Noise Factory model \texttt{LNF\char`_LNC\char`_4\char`_8C}) offering 40 dB of gain and roughly 2K of added noise.  At room temperature the signal is sent through another amplifier (MITEQ AMF-5D-00101200-23-10P-LPN) offering another 40 dB of gain.  All measurements were performed with a Keysight PNA vector network analyzer model N5230C.

The resonance of the conical Ti-6Al-4V sample at 20 mK was found to be $\nicefrac{\omega_0}{2\pi}=7.50$ GHz in agreement with simulation.    To study the temperature dependence of the superconducting microwave surface impedance, we varied the applied heat delivered to the mixing chamber of our dilution refrigerator.  At each temperature, the samples were allowed to equilibrate for two hours before measurements were taken.  This time scale was determined by a series of overnight runs at a fixed temperature of 475mK and 500mK to determine the time constant for the measured $|S_{21}|^2$ to have only gaussian noise about a fixed mean and no systematic variation.    All fits to the measured $|S_{21}|^2$ response from the vector network analyzer were performed using a complex coupling quality factor, a resonance frequency, and an internal quality factor as in Ref. \cite{geerlings2012}.

\begin{figure}
\centering
\includegraphics[scale=.2]{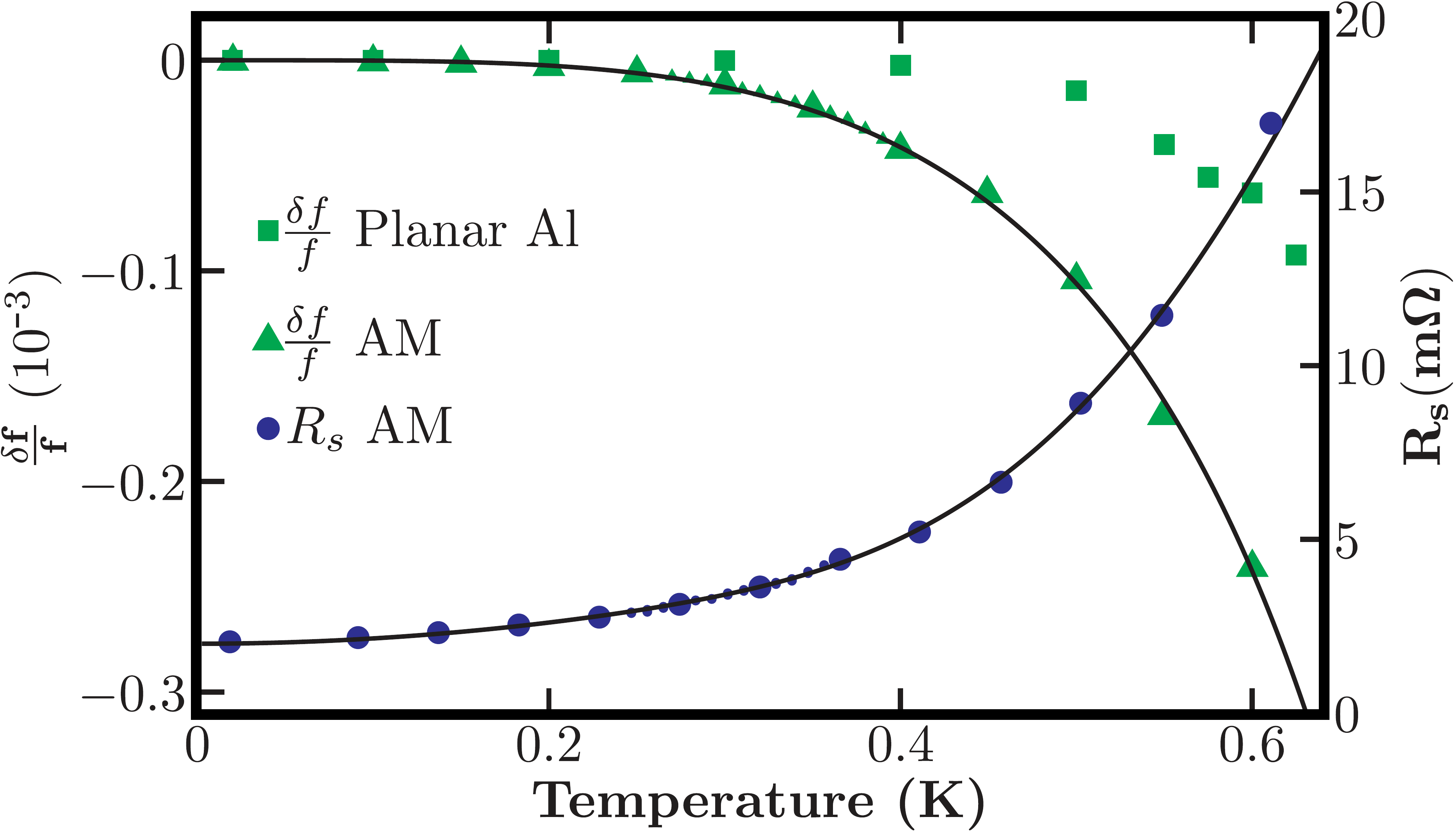}
\caption{Temperature dependence of the superconducting microwave surface impedance. Green triangles show the fractional frequency shift of the Ti-6Al-4V conical resonator as a function of equilibrium temperature.  Shown on the same scale for reference is the fractional frequency shift as a function of temperature (green squares) of a coplanar waveguide aluminum resonator.   Blue dots show the resistance
of the Ti-6Al-4V resonator as a function of temperature.}
{\label{fig:zs}}
\end{figure}

The temperature dependence of the fractional frequency shift ($\frac{\delta f}{f}$) can be related in a straightforward way to the surface reactance of the resonator \cite{gao2008}.  In Fig. \ref{fig:zs}, the fractional frequency shift (green triangles, left y-axis) is plotted as a function of equilibrium temperature recorded on the mixing chamber.   Along with the results for the Ti-6Al-4V resonator we show frequency shift results for a simultaneously measured planar aluminum resonator (identical to Ref. \cite{sage2011}).    All points in the figure represent the average of three measurements of $|S_{21}|^2$ and the size of the points are substantially larger than the associated error bars.
The fractional frequency shift temperature dependence is well fit by a simple two fluid model \cite{turneaure1991}:
\begin{eqnarray}
\frac{\delta f(T)}{f} = \frac{-\alpha _{KI}}{2(1-(\frac{T}{T_c})^4)}+\frac{\alpha_{KI}}{2}
\end{eqnarray}
Where $\alpha_{KI}$ is the zero temperature kinetic inductance fraction and $T_c$ is the superconducting transition temperature.  From this fit, we extract a kinetic inductance fraction of $\alpha_{KI}=4\times 10^{\textrm{-}3} \pm2\times 10^{\textrm{-}3}$ and $T_c = 0.95 \pm 0.1 K$.   We note that this fitted value is in very good agreement with the low temperature transition observed in the heat capacity data.  

We relate the extracted kinetic inductance fraction of the resonator to the zero temperature London penetration depth, $\lambda_0$, via the magnetic field, $H$, participation ratio, $p_{mag}$ as \cite{reagor2013}:
\begin{eqnarray}
\label{alphaKI}
\alpha_{KI} = \lambda _0 p_{mag}=\lambda _0 \frac{ \int H^2dA}{\int H^2 dV}.
\end{eqnarray}
Evaluation of (\ref{alphaKI}) requires a calculation of the magnetic field energy stored on the surface, $\int H^2dA$, compared to the magnetic field energy stored in the volume, $\int H^2dV$.  This can be trivially calculated in an electrodynamic solver such as ANSYS HFSS or COMSOL.  For this manuscript, both software suites were used and agreed upon a value of 528 m$^{\textrm{-}1}$ for the conical geometry. This is well within a factor of two of what one can calculate via closed form analytic solutions of a conventional coaxial geometry.  Using this value of the magnetic participation ratio, we extract $\lambda_0=8\pm3~\mu m$.  This length scale is quite large when compared with aluminum (52 nm) \cite{reagor2013}, niobium (32 nm) \cite{turneaure1991}, or even titanium nitride (575 nm) \cite{vissers2010}.  
This means that a tremendous amount of energy is stored in the motion of the Cooper pairs and opens the possibility of a new and interesting material for MKIDs and other applications designed to exploit high kinetic inductance.

Finally, we investigate the resistive part of the superconducting microwave surface impedance by transforming the extracted internal quality factor ($Q_i\approx 3\times 10^4$ in the quantum regime).  This is accomplished by relating the measured internal quality factor at various temperatures, $Q_i(T)$, to the temperature dependent surface resistance, $R_s(T)$, through the magnetic participation ratio, the permeability of free space, $\mu_0$, and the resonance frequency in angular units, $\omega_0$, as \cite{padamsee2001}:
\begin{eqnarray}
R_s(T) = \frac{\mu_0 \omega_0}{p_{mag}Q_i(T)}.
\end{eqnarray}

In Fig. \ref{fig:zs}, on the right in blue dots, we plot the surface resistance as a function of equilibrium temperature on the mixing chamber.  We fit the measured surface resistance with a functional form that is comprised of a BCS component (exponential), a metallic, Fermi liquid,  component (quadratic), and a residual resistance (constant):
\begin{equation}
R_s(T) = \frac{A}{T}exp(-\frac{\Delta _0}{k_bT}) + BT^2+ R_0.
\end{equation}

At high temperatures, the dominant contribution to the surface resistance is from quasiparticle loss (exponential) and the superconducting gap ($\Delta_0$) is consistent with the fractional frequency shift.  However, at lower temperatures the temperature dependence is dominated by a $T^2$ behavior consistent with a classical Fermi fluid contribution \cite{grimvall1981}.  Power dependent measurements of 
 this resonator displayed no appreciable power dependance in the internal quality factor indicating that dissipation due to low energy defects was not a dominant source of loss in this system\cite{gao2008}.  Given the explicit $T^2$ dependence at low temperature and the lack of a strong power dependence, we attribute the low temperature residual resistance to conductive losses in the normal metal part of the microwave surface impedance. 

In conclusion, we present the first measurement of the London penetration length of one of the most abundant titanium alloys (Ti-6Al-4V) of $8\pm3~\mu$m.  Through careful temperature dependence studies, we extract the effective transition temperature governing the surface impedance at microwave frequencies to find a superconducting gap consistent with heat capacity measurements of the same sample.  Unique to this work, we demonstrate proof of principle additive manufacturing based fabrication and geometry optimization of superconducting resonators which could not be constructed out of one contiguous piece by conventional fabrication techniques.  This work shows that Ti-6Al-4V stands as an unusually high kinetic inductance material and has potential applications in quantum limited amplification in the microwave regime, as well as a single photon detector in the terahertz to gamma ray frequencies.

We would like to thank Scott McCall for assistance with the PPMS system, Sharon Torres for surface characterization, Will Oliver for providing the aluminum planar resonators, and Sean Durham and Jesse Hamblen for solving all our problems yesterday.  This work was performed under the auspices of the U.S. Department of Energy by Lawrence Livermore National Laboratory under Contract DE-AC52-07NA27344.  This work was supported by the Laboratory Directed Research and Development grant 16-SI-004.  LLNL-JRNL-733239.
\bibliographystyle{apsrev4-1}
\bibliography{holland_citations}

\begin{thebibliography}{29}%
\makeatletter
\providecommand \@ifxundefined [1]{%
 \@ifx{#1\undefined}
}%
\providecommand \@ifnum [1]{%
 \ifnum #1\expandafter \@firstoftwo
 \else \expandafter \@secondoftwo
 \fi
}%
\providecommand \@ifx [1]{%
 \ifx #1\expandafter \@firstoftwo
 \else \expandafter \@secondoftwo
 \fi
}%
\providecommand \natexlab [1]{#1}%
\providecommand \enquote  [1]{``#1''}%
\providecommand \bibnamefont  [1]{#1}%
\providecommand \bibfnamefont [1]{#1}%
\providecommand \citenamefont [1]{#1}%
\providecommand \href@noop [0]{\@secondoftwo}%
\providecommand \href [0]{\begingroup \@sanitize@url \@href}%
\providecommand \@href[1]{\@@startlink{#1}\@@href}%
\providecommand \@@href[1]{\endgroup#1\@@endlink}%
\providecommand \@sanitize@url [0]{\catcode `\\12\catcode `\$12\catcode
  `\&12\catcode `\#12\catcode `\^12\catcode `\_12\catcode `\%12\relax}%
\providecommand \@@startlink[1]{}%
\providecommand \@@endlink[0]{}%
\providecommand \url  [0]{\begingroup\@sanitize@url \@url }%
\providecommand \@url [1]{\endgroup\@href {#1}{\urlprefix }}%
\providecommand \urlprefix  [0]{URL }%
\providecommand \Eprint [0]{\href }%
\providecommand \doibase [0]{http://dx.doi.org/}%
\providecommand \selectlanguage [0]{\@gobble}%
\providecommand \bibinfo  [0]{\@secondoftwo}%
\providecommand \bibfield  [0]{\@secondoftwo}%
\providecommand \translation [1]{[#1]}%
\providecommand \BibitemOpen [0]{}%
\providecommand \bibitemStop [0]{}%
\providecommand \bibitemNoStop [0]{.\EOS\space}%
\providecommand \EOS [0]{\spacefactor3000\relax}%
\providecommand \BibitemShut  [1]{\csname bibitem#1\endcsname}%
\let\auto@bib@innerbib\@empty
\bibitem [{\citenamefont {Landau}\ and\ \citenamefont
  {Ginzburg}(1950)}]{landau1950}%
  \BibitemOpen
  \bibfield  {author} {\bibinfo {author} {\bibfnamefont {L.~D.}\ \bibnamefont
  {Landau}}\ and\ \bibinfo {author} {\bibfnamefont {V.}~\bibnamefont
  {Ginzburg}},\ }\href@noop {} {\bibfield  {journal} {\bibinfo  {journal} {Zh.
  Eksp. Teor. Fiz.}\ }\textbf {\bibinfo {volume} {20}},\ \bibinfo {pages}
  {1064} (\bibinfo {year} {1950})}\BibitemShut {NoStop}%
\bibitem [{\citenamefont {Gor'kov}(1959)}]{gor1959}%
  \BibitemOpen
  \bibfield  {author} {\bibinfo {author} {\bibfnamefont {L.~P.}\ \bibnamefont
  {Gor'kov}},\ }\href@noop {} {\bibfield  {journal} {\bibinfo  {journal} {Sov.
  Phys. JETP}\ }\textbf {\bibinfo {volume} {9}},\ \bibinfo {pages} {1364}
  (\bibinfo {year} {1959})}\BibitemShut {NoStop}%
\bibitem [{\citenamefont {Tinkham}(1996)}]{tinkham1996}%
  \BibitemOpen
  \bibfield  {author} {\bibinfo {author} {\bibfnamefont {M.}~\bibnamefont
  {Tinkham}},\ }\href@noop {} {\emph {\bibinfo {title} {Introduction to
  superconductivity}}}\ (\bibinfo  {publisher} {Courier Corporation},\ \bibinfo
  {year} {1996})\BibitemShut {NoStop}%
\bibitem [{\citenamefont {Day}\ \emph {et~al.}(2003)\citenamefont {Day},
  \citenamefont {LeDuc}, \citenamefont {Mazin}, \citenamefont {Vayonakis},\
  and\ \citenamefont {Zmuidzinas}}]{day2003}%
  \BibitemOpen
  \bibfield  {author} {\bibinfo {author} {\bibfnamefont {P.~K.}\ \bibnamefont
  {Day}}, \bibinfo {author} {\bibfnamefont {H.~G.}\ \bibnamefont {LeDuc}},
  \bibinfo {author} {\bibfnamefont {B.~A.}\ \bibnamefont {Mazin}}, \bibinfo
  {author} {\bibfnamefont {A.}~\bibnamefont {Vayonakis}}, \ and\ \bibinfo
  {author} {\bibfnamefont {J.}~\bibnamefont {Zmuidzinas}},\ }\href@noop {}
  {\bibfield  {journal} {\bibinfo  {journal} {Nature}\ }\textbf {\bibinfo
  {volume} {425}},\ \bibinfo {pages} {817} (\bibinfo {year}
  {2003})}\BibitemShut {NoStop}%
\bibitem [{\citenamefont {Leyens}\ and\ \citenamefont
  {Peters}(2003)}]{leyens2003}%
  \BibitemOpen
  \bibfield  {author} {\bibinfo {author} {\bibfnamefont {C.}~\bibnamefont
  {Leyens}}\ and\ \bibinfo {author} {\bibfnamefont {M.}~\bibnamefont
  {Peters}},\ }\href@noop {} {\emph {\bibinfo {title} {Titanium and titanium
  alloys: fundamentals and applications}}}\ (\bibinfo  {publisher} {John Wiley
  \&amp; Sons},\ \bibinfo {year} {2003})\BibitemShut {NoStop}%
\bibitem [{\citenamefont {Creedon}\ \emph {et~al.}(2016)\citenamefont
  {Creedon}, \citenamefont {Goryachev}, \citenamefont {Kostylev}, \citenamefont
  {Sercombe},\ and\ \citenamefont {Tobar}}]{creedon2016}%
  \BibitemOpen
  \bibfield  {author} {\bibinfo {author} {\bibfnamefont {D.~L.}\ \bibnamefont
  {Creedon}}, \bibinfo {author} {\bibfnamefont {M.}~\bibnamefont {Goryachev}},
  \bibinfo {author} {\bibfnamefont {N.}~\bibnamefont {Kostylev}}, \bibinfo
  {author} {\bibfnamefont {T.~B.}\ \bibnamefont {Sercombe}}, \ and\ \bibinfo
  {author} {\bibfnamefont {M.~E.}\ \bibnamefont {Tobar}},\ }\href@noop {}
  {\bibfield  {journal} {\bibinfo  {journal} {Appl. Phys. Lett.}\ }\textbf
  {\bibinfo {volume} {109}},\ \bibinfo {pages} {032601} (\bibinfo {year}
  {2016})}\BibitemShut {NoStop}%
\bibitem [{\citenamefont {Gorman}\ \emph {et~al.}(2001)\citenamefont {Gorman},
  \citenamefont {Woodfield},\ and\ \citenamefont {Link}}]{gorman2001}%
  \BibitemOpen
  \bibfield  {author} {\bibinfo {author} {\bibfnamefont {M.~D.}\ \bibnamefont
  {Gorman}}, \bibinfo {author} {\bibfnamefont {A.~P.}\ \bibnamefont
  {Woodfield}}, \ and\ \bibinfo {author} {\bibfnamefont {B.~A.}\ \bibnamefont
  {Link}},\ }\href@noop {} {\enquote {\bibinfo {title} {Heat treatment for
  improved properties of alpha-beta titanium-base alloys},}\ } (\bibinfo {year}
  {2001}),\ \bibinfo {note} {uS Patent 6,284,070}\BibitemShut {NoStop}%
\bibitem [{\citenamefont {Kobryn}\ and\ \citenamefont
  {Semiatin}(2001)}]{kobryn2001}%
  \BibitemOpen
  \bibfield  {author} {\bibinfo {author} {\bibfnamefont {P.}~\bibnamefont
  {Kobryn}}\ and\ \bibinfo {author} {\bibfnamefont {S.}~\bibnamefont
  {Semiatin}},\ }\href@noop {} {\bibfield  {journal} {\bibinfo  {journal} {JOM
  Journal of the Minerals, Metals and Materials Society}\ }\textbf {\bibinfo
  {volume} {53}},\ \bibinfo {pages} {40} (\bibinfo {year} {2001})}\BibitemShut
  {NoStop}%
\bibitem [{\citenamefont {Thijs}\ \emph {et~al.}(2010)\citenamefont {Thijs},
  \citenamefont {Verhaeghe}, \citenamefont {Craeghs}, \citenamefont
  {Van~Humbeeck},\ and\ \citenamefont {Kruth}}]{thijs2010}%
  \BibitemOpen
  \bibfield  {author} {\bibinfo {author} {\bibfnamefont {L.}~\bibnamefont
  {Thijs}}, \bibinfo {author} {\bibfnamefont {F.}~\bibnamefont {Verhaeghe}},
  \bibinfo {author} {\bibfnamefont {T.}~\bibnamefont {Craeghs}}, \bibinfo
  {author} {\bibfnamefont {J.}~\bibnamefont {Van~Humbeeck}}, \ and\ \bibinfo
  {author} {\bibfnamefont {J.-P.}\ \bibnamefont {Kruth}},\ }\href@noop {}
  {\bibfield  {journal} {\bibinfo  {journal} {Acta Materialia}\ }\textbf
  {\bibinfo {volume} {58}},\ \bibinfo {pages} {3303} (\bibinfo {year}
  {2010})}\BibitemShut {NoStop}%
\bibitem [{\citenamefont {Pyka}\ \emph {et~al.}(2012)\citenamefont {Pyka},
  \citenamefont {Burakowski}, \citenamefont {Kerckhofs}, \citenamefont
  {Moesen}, \citenamefont {Van~Bael}, \citenamefont {Schrooten},\ and\
  \citenamefont {Wevers}}]{pyka2012}%
  \BibitemOpen
  \bibfield  {author} {\bibinfo {author} {\bibfnamefont {G.}~\bibnamefont
  {Pyka}}, \bibinfo {author} {\bibfnamefont {A.}~\bibnamefont {Burakowski}},
  \bibinfo {author} {\bibfnamefont {G.}~\bibnamefont {Kerckhofs}}, \bibinfo
  {author} {\bibfnamefont {M.}~\bibnamefont {Moesen}}, \bibinfo {author}
  {\bibfnamefont {S.}~\bibnamefont {Van~Bael}}, \bibinfo {author}
  {\bibfnamefont {J.}~\bibnamefont {Schrooten}}, \ and\ \bibinfo {author}
  {\bibfnamefont {M.}~\bibnamefont {Wevers}},\ }\href@noop {} {\bibfield
  {journal} {\bibinfo  {journal} {Advanced Engineering Materials}\ }\textbf
  {\bibinfo {volume} {14}},\ \bibinfo {pages} {363} (\bibinfo {year}
  {2012})}\BibitemShut {NoStop}%
\bibitem [{\citenamefont {Kruth}\ \emph {et~al.}(2007)\citenamefont {Kruth},
  \citenamefont {Levy}, \citenamefont {Klocke},\ and\ \citenamefont
  {Childs}}]{kruth2007}%
  \BibitemOpen
  \bibfield  {author} {\bibinfo {author} {\bibfnamefont {J.-P.}\ \bibnamefont
  {Kruth}}, \bibinfo {author} {\bibfnamefont {G.}~\bibnamefont {Levy}},
  \bibinfo {author} {\bibfnamefont {F.}~\bibnamefont {Klocke}}, \ and\ \bibinfo
  {author} {\bibfnamefont {T.}~\bibnamefont {Childs}},\ }\href@noop {}
  {\bibfield  {journal} {\bibinfo  {journal} {CIRP Annals-Manufacturing
  Technology}\ }\textbf {\bibinfo {volume} {56}},\ \bibinfo {pages} {730}
  (\bibinfo {year} {2007})}\BibitemShut {NoStop}%
\bibitem [{\citenamefont {Matthews}\ \emph {et~al.}(2016)\citenamefont
  {Matthews}, \citenamefont {Guss}, \citenamefont {Khairallah}, \citenamefont
  {Rubenchik}, \citenamefont {Depond},\ and\ \citenamefont
  {King}}]{matthews2016}%
  \BibitemOpen
  \bibfield  {author} {\bibinfo {author} {\bibfnamefont {M.~J.}\ \bibnamefont
  {Matthews}}, \bibinfo {author} {\bibfnamefont {G.}~\bibnamefont {Guss}},
  \bibinfo {author} {\bibfnamefont {S.~A.}\ \bibnamefont {Khairallah}},
  \bibinfo {author} {\bibfnamefont {A.~M.}\ \bibnamefont {Rubenchik}}, \bibinfo
  {author} {\bibfnamefont {P.~J.}\ \bibnamefont {Depond}}, \ and\ \bibinfo
  {author} {\bibfnamefont {W.~E.}\ \bibnamefont {King}},\ }\href@noop {}
  {\bibfield  {journal} {\bibinfo  {journal} {Acta Materialia}\ }\textbf
  {\bibinfo {volume} {114}},\ \bibinfo {pages} {33} (\bibinfo {year}
  {2016})}\BibitemShut {NoStop}%
\bibitem [{\citenamefont {Clark}\ \emph {et~al.}(1970)\citenamefont {Clark},
  \citenamefont {Childs},\ and\ \citenamefont {Wallace}}]{clark1970}%
  \BibitemOpen
  \bibfield  {author} {\bibinfo {author} {\bibfnamefont {A.}~\bibnamefont
  {Clark}}, \bibinfo {author} {\bibfnamefont {G.}~\bibnamefont {Childs}}, \
  and\ \bibinfo {author} {\bibfnamefont {G.}~\bibnamefont {Wallace}},\
  }\href@noop {} {\bibfield  {journal} {\bibinfo  {journal} {Cryogenics}\
  }\textbf {\bibinfo {volume} {10}},\ \bibinfo {pages} {295} (\bibinfo {year}
  {1970})}\BibitemShut {NoStop}%
\bibitem [{\citenamefont {Lepper}\ \emph {et~al.}(1972)\citenamefont {Lepper},
  \citenamefont {Wolff},\ and\ \citenamefont {Mills}}]{lepper1972}%
  \BibitemOpen
  \bibfield  {author} {\bibinfo {author} {\bibfnamefont {R.}~\bibnamefont
  {Lepper}}, \bibinfo {author} {\bibfnamefont {E.~G.}\ \bibnamefont {Wolff}}, \
  and\ \bibinfo {author} {\bibfnamefont {G.~J.}\ \bibnamefont {Mills}},\
  }\href@noop {} {\emph {\bibinfo {title} {Ac permeability studies of ternary
  alloys at cryogenic temperatures}}},\ \bibinfo {type} {Tech. Rep.}\ (\bibinfo
   {institution} {Northrop Corp., Hawthorne, CA},\ \bibinfo {year}
  {1972})\BibitemShut {NoStop}%
\bibitem [{\citenamefont {Wolff}\ \emph {et~al.}(1973)\citenamefont {Wolff},
  \citenamefont {Lepper},\ and\ \citenamefont {Mills}}]{wolff1973}%
  \BibitemOpen
  \bibfield  {author} {\bibinfo {author} {\bibfnamefont {E.}~\bibnamefont
  {Wolff}}, \bibinfo {author} {\bibfnamefont {R.}~\bibnamefont {Lepper}}, \
  and\ \bibinfo {author} {\bibfnamefont {G.}~\bibnamefont {Mills}},\
  }\href@noop {} {\bibfield  {journal} {\bibinfo  {journal} {Titanium Science
  and Technology,}\ }\textbf {\bibinfo {volume} {2}} (\bibinfo {year}
  {1973})}\BibitemShut {NoStop}%
\bibitem [{\citenamefont {Umezawa}\ and\ \citenamefont
  {Ishikawa}(1992)}]{umezawa1992}%
  \BibitemOpen
  \bibfield  {author} {\bibinfo {author} {\bibfnamefont {O.}~\bibnamefont
  {Umezawa}}\ and\ \bibinfo {author} {\bibfnamefont {K.}~\bibnamefont
  {Ishikawa}},\ }\href@noop {} {\bibfield  {journal} {\bibinfo  {journal}
  {Cryogenics}\ }\textbf {\bibinfo {volume} {32}},\ \bibinfo {pages} {873}
  (\bibinfo {year} {1992})}\BibitemShut {NoStop}%
\bibitem [{\citenamefont {Divakar}\ \emph {et~al.}(2008)\citenamefont
  {Divakar}, \citenamefont {Henry}, \citenamefont {Kraus},\ and\ \citenamefont
  {Tolhurst}}]{divakar2008}%
  \BibitemOpen
  \bibfield  {author} {\bibinfo {author} {\bibfnamefont {U.}~\bibnamefont
  {Divakar}}, \bibinfo {author} {\bibfnamefont {S.}~\bibnamefont {Henry}},
  \bibinfo {author} {\bibfnamefont {H.}~\bibnamefont {Kraus}}, \ and\ \bibinfo
  {author} {\bibfnamefont {A.}~\bibnamefont {Tolhurst}},\ }\href@noop {}
  {\bibfield  {journal} {\bibinfo  {journal} {Superconductor Science and
  Technology}\ }\textbf {\bibinfo {volume} {21}},\ \bibinfo {pages} {065021}
  (\bibinfo {year} {2008})}\BibitemShut {NoStop}%
\bibitem [{\citenamefont {Divakar}\ \emph {et~al.}(2010)\citenamefont
  {Divakar}, \citenamefont {Henry}, \citenamefont {Kraus},\ and\ \citenamefont
  {Tolhurst}}]{divakar2010}%
  \BibitemOpen
  \bibfield  {author} {\bibinfo {author} {\bibfnamefont {U.}~\bibnamefont
  {Divakar}}, \bibinfo {author} {\bibfnamefont {S.}~\bibnamefont {Henry}},
  \bibinfo {author} {\bibfnamefont {H.}~\bibnamefont {Kraus}}, \ and\ \bibinfo
  {author} {\bibfnamefont {A.}~\bibnamefont {Tolhurst}},\ }\href@noop {}
  {\bibfield  {journal} {\bibinfo  {journal} {Superconductor Science and
  Technology}\ }\textbf {\bibinfo {volume} {23}},\ \bibinfo {pages} {129801}
  (\bibinfo {year} {2010})}\BibitemShut {NoStop}%
\bibitem [{\citenamefont {Ridgeon}\ \emph {et~al.}(2016)\citenamefont
  {Ridgeon}, \citenamefont {Raine}, \citenamefont {Halliday}, \citenamefont
  {Lakrimi}, \citenamefont {Thomas},\ and\ \citenamefont
  {Hampshire}}]{ridgeon2016}%
  \BibitemOpen
  \bibfield  {author} {\bibinfo {author} {\bibfnamefont {F.}~\bibnamefont
  {Ridgeon}}, \bibinfo {author} {\bibfnamefont {M.}~\bibnamefont {Raine}},
  \bibinfo {author} {\bibfnamefont {D.}~\bibnamefont {Halliday}}, \bibinfo
  {author} {\bibfnamefont {M.}~\bibnamefont {Lakrimi}}, \bibinfo {author}
  {\bibfnamefont {A.}~\bibnamefont {Thomas}}, \ and\ \bibinfo {author}
  {\bibfnamefont {D.}~\bibnamefont {Hampshire}},\ }\href@noop {} {\bibfield
  {journal} {\bibinfo  {journal} {IEEE Transactions on Applied
  Superconductivity}\ } (\bibinfo {year} {2016})}\BibitemShut {NoStop}%
\bibitem [{\citenamefont {Reagor}\ \emph {et~al.}(2016)\citenamefont {Reagor},
  \citenamefont {Pfaff}, \citenamefont {Axline}, \citenamefont {Heeres},
  \citenamefont {Ofek}, \citenamefont {Sliwa}, \citenamefont {Holland},
  \citenamefont {Wang}, \citenamefont {Blumoff}, \citenamefont {Chou} \emph
  {et~al.}}]{reagor2016}%
  \BibitemOpen
  \bibfield  {author} {\bibinfo {author} {\bibfnamefont {M.}~\bibnamefont
  {Reagor}}, \bibinfo {author} {\bibfnamefont {W.}~\bibnamefont {Pfaff}},
  \bibinfo {author} {\bibfnamefont {C.}~\bibnamefont {Axline}}, \bibinfo
  {author} {\bibfnamefont {R.~W.}\ \bibnamefont {Heeres}}, \bibinfo {author}
  {\bibfnamefont {N.}~\bibnamefont {Ofek}}, \bibinfo {author} {\bibfnamefont
  {K.}~\bibnamefont {Sliwa}}, \bibinfo {author} {\bibfnamefont {E.~T.}\
  \bibnamefont {Holland}}, \bibinfo {author} {\bibfnamefont {C.}~\bibnamefont
  {Wang}}, \bibinfo {author} {\bibfnamefont {J.}~\bibnamefont {Blumoff}},
  \bibinfo {author} {\bibfnamefont {K.}~\bibnamefont {Chou}},  \emph {et~al.},\
  }\href@noop {} {\bibfield  {journal} {\bibinfo  {journal} {Physical Review
  B}\ }\textbf {\bibinfo {volume} {94}},\ \bibinfo {pages} {014506} (\bibinfo
  {year} {2016})}\BibitemShut {NoStop}%
\bibitem [{\citenamefont {Wang}\ \emph {et~al.}(2016)\citenamefont {Wang},
  \citenamefont {Gao}, \citenamefont {Reinhold}, \citenamefont {Heeres},
  \citenamefont {Ofek}, \citenamefont {Chou}, \citenamefont {Axline},
  \citenamefont {Reagor}, \citenamefont {Blumoff}, \citenamefont {Sliwa} \emph
  {et~al.}}]{wang2016}%
  \BibitemOpen
  \bibfield  {author} {\bibinfo {author} {\bibfnamefont {C.}~\bibnamefont
  {Wang}}, \bibinfo {author} {\bibfnamefont {Y.~Y.}\ \bibnamefont {Gao}},
  \bibinfo {author} {\bibfnamefont {P.}~\bibnamefont {Reinhold}}, \bibinfo
  {author} {\bibfnamefont {R.}~\bibnamefont {Heeres}}, \bibinfo {author}
  {\bibfnamefont {N.}~\bibnamefont {Ofek}}, \bibinfo {author} {\bibfnamefont
  {K.}~\bibnamefont {Chou}}, \bibinfo {author} {\bibfnamefont {C.}~\bibnamefont
  {Axline}}, \bibinfo {author} {\bibfnamefont {M.}~\bibnamefont {Reagor}},
  \bibinfo {author} {\bibfnamefont {J.}~\bibnamefont {Blumoff}}, \bibinfo
  {author} {\bibfnamefont {K.}~\bibnamefont {Sliwa}},  \emph {et~al.},\
  }\href@noop {} {\bibfield  {journal} {\bibinfo  {journal} {Science}\ }\textbf
  {\bibinfo {volume} {352}},\ \bibinfo {pages} {1087} (\bibinfo {year}
  {2016})}\BibitemShut {NoStop}%
\bibitem [{\citenamefont {Reagor}\ \emph {et~al.}(2013)\citenamefont {Reagor},
  \citenamefont {Paik}, \citenamefont {Catelani}, \citenamefont {Sun},
  \citenamefont {Axline}, \citenamefont {Holland}, \citenamefont {Pop},
  \citenamefont {Masluk}, \citenamefont {Brecht}, \citenamefont {Frunzio} \emph
  {et~al.}}]{reagor2013}%
  \BibitemOpen
  \bibfield  {author} {\bibinfo {author} {\bibfnamefont {M.~J.}\ \bibnamefont
  {Reagor}}, \bibinfo {author} {\bibfnamefont {H.}~\bibnamefont {Paik}},
  \bibinfo {author} {\bibfnamefont {G.}~\bibnamefont {Catelani}}, \bibinfo
  {author} {\bibfnamefont {L.}~\bibnamefont {Sun}}, \bibinfo {author}
  {\bibfnamefont {C.}~\bibnamefont {Axline}}, \bibinfo {author} {\bibfnamefont
  {E.~T.}\ \bibnamefont {Holland}}, \bibinfo {author} {\bibfnamefont {I.~M.}\
  \bibnamefont {Pop}}, \bibinfo {author} {\bibfnamefont {N.~A.}\ \bibnamefont
  {Masluk}}, \bibinfo {author} {\bibfnamefont {T.}~\bibnamefont {Brecht}},
  \bibinfo {author} {\bibfnamefont {L.}~\bibnamefont {Frunzio}},  \emph
  {et~al.},\ }\href@noop {} {\bibfield  {journal} {\bibinfo  {journal} {Appl.
  Phys. Lett.}\ }\textbf {\bibinfo {volume} {102}},\ \bibinfo {pages} {192604}
  (\bibinfo {year} {2013})}\BibitemShut {NoStop}%
\bibitem [{\citenamefont {Geerlings}\ \emph {et~al.}(2012)\citenamefont
  {Geerlings}, \citenamefont {Shankar}, \citenamefont {Edwards}, \citenamefont
  {Frunzio}, \citenamefont {Schoelkopf},\ and\ \citenamefont
  {Devoret}}]{geerlings2012}%
  \BibitemOpen
  \bibfield  {author} {\bibinfo {author} {\bibfnamefont {K.}~\bibnamefont
  {Geerlings}}, \bibinfo {author} {\bibfnamefont {S.}~\bibnamefont {Shankar}},
  \bibinfo {author} {\bibfnamefont {E.}~\bibnamefont {Edwards}}, \bibinfo
  {author} {\bibfnamefont {L.}~\bibnamefont {Frunzio}}, \bibinfo {author}
  {\bibfnamefont {R.}~\bibnamefont {Schoelkopf}}, \ and\ \bibinfo {author}
  {\bibfnamefont {M.}~\bibnamefont {Devoret}},\ }\href@noop {} {\bibfield
  {journal} {\bibinfo  {journal} {Appl. Phys. Lett.}\ }\textbf {\bibinfo
  {volume} {100}},\ \bibinfo {pages} {192601} (\bibinfo {year}
  {2012})}\BibitemShut {NoStop}%
\bibitem [{\citenamefont {Gao}\ \emph {et~al.}(2008)\citenamefont {Gao},
  \citenamefont {Daal}, \citenamefont {Vayonakis}, \citenamefont {Kumar},
  \citenamefont {Zmuidzinas}, \citenamefont {Sadoulet}, \citenamefont {Mazin},
  \citenamefont {Day},\ and\ \citenamefont {Leduc}}]{gao2008}%
  \BibitemOpen
  \bibfield  {author} {\bibinfo {author} {\bibfnamefont {J.}~\bibnamefont
  {Gao}}, \bibinfo {author} {\bibfnamefont {M.}~\bibnamefont {Daal}}, \bibinfo
  {author} {\bibfnamefont {A.}~\bibnamefont {Vayonakis}}, \bibinfo {author}
  {\bibfnamefont {S.}~\bibnamefont {Kumar}}, \bibinfo {author} {\bibfnamefont
  {J.}~\bibnamefont {Zmuidzinas}}, \bibinfo {author} {\bibfnamefont
  {B.}~\bibnamefont {Sadoulet}}, \bibinfo {author} {\bibfnamefont {B.~A.}\
  \bibnamefont {Mazin}}, \bibinfo {author} {\bibfnamefont {P.~K.}\ \bibnamefont
  {Day}}, \ and\ \bibinfo {author} {\bibfnamefont {H.~G.}\ \bibnamefont
  {Leduc}},\ }\href@noop {} {\bibfield  {journal} {\bibinfo  {journal} {Appl.
  Phys. Lett.}\ }\textbf {\bibinfo {volume} {92}},\ \bibinfo {pages} {152505}
  (\bibinfo {year} {2008})}\BibitemShut {NoStop}%
\bibitem [{\citenamefont {Sage}\ \emph {et~al.}(2011)\citenamefont {Sage},
  \citenamefont {Bolkhovsky}, \citenamefont {Oliver}, \citenamefont {Turek},\
  and\ \citenamefont {Welander}}]{sage2011}%
  \BibitemOpen
  \bibfield  {author} {\bibinfo {author} {\bibfnamefont {J.~M.}\ \bibnamefont
  {Sage}}, \bibinfo {author} {\bibfnamefont {V.}~\bibnamefont {Bolkhovsky}},
  \bibinfo {author} {\bibfnamefont {W.~D.}\ \bibnamefont {Oliver}}, \bibinfo
  {author} {\bibfnamefont {B.}~\bibnamefont {Turek}}, \ and\ \bibinfo {author}
  {\bibfnamefont {P.~B.}\ \bibnamefont {Welander}},\ }\href@noop {} {\bibfield
  {journal} {\bibinfo  {journal} {J. Appl. Phys.}\ }\textbf {\bibinfo {volume}
  {109}},\ \bibinfo {pages} {063915} (\bibinfo {year} {2011})}\BibitemShut
  {NoStop}%
\bibitem [{\citenamefont {Turneaure}\ \emph {et~al.}(1991)\citenamefont
  {Turneaure}, \citenamefont {Halbritter},\ and\ \citenamefont
  {Schwettman}}]{turneaure1991}%
  \BibitemOpen
  \bibfield  {author} {\bibinfo {author} {\bibfnamefont {J.}~\bibnamefont
  {Turneaure}}, \bibinfo {author} {\bibfnamefont {J.}~\bibnamefont
  {Halbritter}}, \ and\ \bibinfo {author} {\bibfnamefont {H.}~\bibnamefont
  {Schwettman}},\ }\href@noop {} {\bibfield  {journal} {\bibinfo  {journal}
  {Journal of Superconductivity}\ }\textbf {\bibinfo {volume} {4}},\ \bibinfo
  {pages} {341} (\bibinfo {year} {1991})}\BibitemShut {NoStop}%
\bibitem [{\citenamefont {Vissers}\ \emph {et~al.}(2010)\citenamefont
  {Vissers}, \citenamefont {Gao}, \citenamefont {Wisbey}, \citenamefont {Hite},
  \citenamefont {Tsuei}, \citenamefont {Corcoles}, \citenamefont {Steffen},\
  and\ \citenamefont {Pappas}}]{vissers2010}%
  \BibitemOpen
  \bibfield  {author} {\bibinfo {author} {\bibfnamefont {M.~R.}\ \bibnamefont
  {Vissers}}, \bibinfo {author} {\bibfnamefont {J.}~\bibnamefont {Gao}},
  \bibinfo {author} {\bibfnamefont {D.~S.}\ \bibnamefont {Wisbey}}, \bibinfo
  {author} {\bibfnamefont {D.~A.}\ \bibnamefont {Hite}}, \bibinfo {author}
  {\bibfnamefont {C.~C.}\ \bibnamefont {Tsuei}}, \bibinfo {author}
  {\bibfnamefont {A.~D.}\ \bibnamefont {Corcoles}}, \bibinfo {author}
  {\bibfnamefont {M.}~\bibnamefont {Steffen}}, \ and\ \bibinfo {author}
  {\bibfnamefont {D.~P.}\ \bibnamefont {Pappas}},\ }\href@noop {} {\bibfield
  {journal} {\bibinfo  {journal} {Applied Physics Letters}\ }\textbf {\bibinfo
  {volume} {97}},\ \bibinfo {pages} {232509} (\bibinfo {year}
  {2010})}\BibitemShut {NoStop}%
\bibitem [{\citenamefont {Padamsee}(2001)}]{padamsee2001}%
  \BibitemOpen
  \bibfield  {author} {\bibinfo {author} {\bibfnamefont {H.}~\bibnamefont
  {Padamsee}},\ }\href@noop {} {\bibfield  {journal} {\bibinfo  {journal}
  {Supercond. Sci. Technol.}\ }\textbf {\bibinfo {volume} {14}},\ \bibinfo
  {pages} {R28} (\bibinfo {year} {2001})}\BibitemShut {NoStop}%
\bibitem [{\citenamefont {Grimvall}(1981)}]{grimvall1981}%
  \BibitemOpen
  \bibfield  {author} {\bibinfo {author} {\bibfnamefont {G.}~\bibnamefont
  {Grimvall}},\ }\href@noop {} {\emph {\bibinfo {title} {The electron-phonon
  interaction in metals}}},\ Vol.~\bibinfo {volume} {8}\ (\bibinfo  {publisher}
  {North-Holland Amsterdam},\ \bibinfo {year} {1981})\BibitemShut {NoStop}%
\end{thebibliography}%
\end{document}